# Structural study of α-Bi$_2$O$_3$ under pressure

A. L. J. Pereira[1,*], D. Errandonea[2], A. Beltrán[3], L. Gracia[3], O. Gomis[4], J. A. Sans[1], B. Garcia-Domene[2], A. Miquel-Veyrat[2], F. J. Manjón[1], A. Muñoz[5], and C. Popescu[6]

[1]*Instituto de Diseño para la Fabricación y Producción Automatizada, MALTA Consolider Team, Universitat Politècnica de Valencia, 46022 València, Spain*
[2]*Departamento de Física Aplicada-ICMUV, MALTA Consolider Team, Universidad de Valencia, Edificio de Investigación, C/Dr. Moliner 50, Burjassot, 46100 Valencia, Spain*
[3]*Departament de Química Física i Analítica, MALTA Consolider Team, Universitat Jaume I, 12071 Castelló, Spain*
[4]*Centro de Tecnologías Físicas, MALTA Consolider Team, Universitat Politècnica de Valencia, 46022 València, Spain*
[5]*Departamento de Física Fundamental II, Universidad de La Laguna, La Laguna, E-38205 Tenerife, Spain*
[6]*CELLS-ALBA Synchrotron Light Facility, Cerdanyola, 08290 Barcelona, Spain*

* Corresponding author: andeje@upvnet.upv.es

**Abstract.** An experimental and theoretical study of the structural properties of monoclinic bismuth oxide (α-Bi$_2$O$_3$) under high pressures is here reported. Both synthetic and mineral bismite powder samples have been compressed up to 45 GPa and their equations of state have been determined with angle-dispersive x-ray diffraction measurements. Experimental results have been also compared to theoretical calculations which suggest the possibility of several phase transitions below 10 GPa. However, experiments reveal only a pressure-induced amorphisation between 15 and 25 GPa, depending on sample quality and deviatoric stresses. The amorphous phase has been followed up to 45 GPa and its nature discussed.

## 1. Introduction

Industrially, the bismuth trioxide (Bi$_2$O$_3$) is the most important compound of bismuth since it is a common starting points for bismuth chemistry. The applicability of Bi$_2$O$_3$ extends from fireworks to oxygen gas sensors and solid oxide fuel cells **[1-6]**. Interest in Bi$_2$O$_3$ is also increasing because it shows similar properties as lead(II) oxide (PbO); namely, the ability to form transparent glasses with a high refractive index useful in optical telecommunication and processing devices **[7,8]** and in ecological *lead-free* glasses for several applications **[9,10]**. Furthermore, there is a recent great interest in the properties of Bi$_2$O$_3$ at high temperatures and high pressures. Under these conditions phase transitions to various polymorphs, which are metastable at ambient conditions, have been observed and whose properties could be interesting for a number of applications [11,12].

The most common polymorph of $Bi_2O_3$ found at ambient conditions is the mineral bismite ($\alpha$-$Bi_2O_3$), which crystallizes in the monoclinic $P2_1/c$, space group (SG) No. 14 [13]. In this phase, the unit cell contains two Bi (Bi-I and Bi-II) atoms located at $4e$ Wyckoff sites and three O (O-I, O-II, and O-III) atoms located at $4e$ Wyckoff sites (see **Figure 1**). The two Bi atoms have different coordination to O atoms: Bi-I has five-fold coordination (two O-I, two O-III, and one O-II) while Bi-II has six-fold coordination (two O-I, two O-III, and two O-II). $Bi_2O_3$ also presents several structures depending on the thermal history. Heating $\alpha$-$Bi_2O_3$ above 730°C results in the formation of $\delta$-$Bi_2O_3$ (SG $Fm$-$3m$, No. 225) with cubic fluorite-type crystal structure. On the other hand, on cooling $\delta$-$Bi_2O_3$ it is possible to form two intermediate metastable phases at ambient conditions: the tetragonal $\beta$ phase (SG $P$-$421c$, No. 114), also known as sphaerobismoite, at ~650 °C, and the body-centered cubic $\gamma$ phase (SG $I23$, No. 197) at ~640 °C [3,13].

Pressure, together with temperature, is a key external variable which determines the structure and properties of solids. The most dramatic effects induced by pressure are structural solid-solid transformations. In this respect, new phases of $Bi_2O_3$ have been recently found on increasing pressure and temperature. Starting with the $\alpha$ phase, Atou *et al.* [14] obtained a hexagonal polymorph with A-type structure (SG $P$-$3m1$, No. 164), typical of rare-earth sesquioxides, after compressing the sample to 6 GPa and heating at 880°C for 30 min. However, the existence of this phase was questioned by Ghedia *et al.* [11], who used a similar procedure of pressurization, heating, and release, but identified two different metastable polymorphs of $Bi_2O_3$ at ambient conditions: HP-$Bi_2O_3$ (SG $P31c$, No.159) and R-$Bi_2O_3$ (SG $P2_1/c$, No.14). HP-$Bi_2O_3$ has a noncentrosymmetric trigonal symmetry and, after some months at room temperature (or after thermal annealing), transforms to the monoclinic R-$Bi_2O_3$ structure. Finally, R-$Bi_2O_3$ transforms to $\alpha$-$Bi_2O_3$.

The metastable HP-$Bi_2O_3$ phase is built from a 3D network of slightly distorted $BiO_6$ polyhedra and strongly distorted $BiO_5$ polyhedra. This phase has been recently studied by x-ray and neutron diffraction at high pressures and it has been found to undergo a *translation gleiche* phase transition at ~2 GPa to a hexagonal structure, named HPC-$Bi_2O_3$ (SG $P63mc$, No. 186), which is stable up to 35 GPa [12]. The HPC phase is a supergroup of the HP phase and is not quenchable at ambient conditions. The HPC phase is built from a 3D network of distorted $BiO_6$ polyhedra and distorted $BiO_7$ polyhedra. The equation of state of both HP and HPC phases also were determined [12]. However, scarce information is known about $\alpha$-$Bi_2O_3$ at high pressures despite its industrial interest. Only a high-pressure Raman study of $\alpha$-$Bi_2O_3$ proposed its amorphisation above 20 GPa [15], and the equation of state (EOS) of synthetic α-$Bi_2O_3$ was recently studied using shock waves [16].

In this work we report a detailed experimental and theoretical study of the structural properties of α-$Bi_2O_3$ under pressure up to 45 GPa. We report the EOS of the monoclinic phase in both synthetic and mineral samples and compare it with that recently obtained [16] and with our theoretical calculations. The purpose of our study is to understand the structural behavior of α-$Bi_2O_3$ at high pressures in order to compare it with that of other V-group sesquioxides, like $As_2O_3$ [17,18] and $Sb_2O_3$ [19,20]. The complexity of the mechanisms involved in the structural transitions of these compounds (involving amorphisation) at high pressure needs for detailed studies of the evolution of the structural parameters in the different phases in all these sesquioxides in order to understand their polymorphism and the range of stability of each polymorph [21].

## 2. Experimental details

Two types of $Bi_2O_3$ samples were used in this study: i) commercial synthetic powder samples with 99.9% purity (Sigma Aldrich), and ii) natural mineral bismite from San Bernardino County, California (USA). The mineral samples were bright yellow microcrystals of bismite extracted from a quartz matrix. The only impurities detectable by electron microprobe analysis were Si, Al, and Fe at 0.4, 0.1, and 0.1 %WT respectively. Three series of experiments were performed: one in the synthetic sample up to 25 GPa using Ar (quasi-hydrostatic conditions) as pressure-transmitting medium (PTM), one in the mineral sample up to 25 GPa using the same PTM, and one in the synthetic sample up to 45 GPa using 16:3:1 methanol-ethanol-water (MEW, less hydrostatic conditions) as PTM. Angle-dispersive x-ray diffraction (ADXRD) experiments were carried out using a Boehler-Almax diamond-anvil cell (DAC) with diamond culets of 280 μm. The pressure chamber was an 80 μm hole drilled on a 40 μm thick pre-indented fingerprint in a tungsten gasket. Special care was taken to occupy only a small fraction on the pressure chamber with the loaded samples to reduce the possibility of sample bridging between the two diamond anvils. Pressure was determined using ruby fluorescence [22], and, after 6.6 GPa, also the EOS of Ar [23,24]. Experiments were performed at the MSPD beamline at ALBA synchrotron facility [25]. This beamline is equipped with Kirkpatrick-Baez mirrors to focus the monochromatic beam and a Rayonix CCD detector with a 165 mm diameter active area. We used a wavelength of 0.4246 Å and the sample-detector distance during the experiment was set to 280 mm. The 2-D diffraction images were integrated with FIT2D software [26]. Structural analysis was performed with PowderCell [27] and GSAS [28,29].

## 3. Theoretical details

First principles total-energy calculations were carried out within the periodic density functional theory (DFT) framework using CRYSTAL09 program package **[30]**. The Kohn-Sham equations have been solved by means of the exchange-correlation functionals in the generalized gradient approximation (GGA) developed for solids by Perdew, Burke, and Ernzerhof (PBESol) **[31]**. Unlike other program packages, the bulk CRYSTAL calculations are periodic in the three dimensions of the space. The O centers have been described by standard Gaussian basis sets, whereas for the Bi centers the core electrons were described by non-relativistic effective core pseudo-potential [PS] and the valence electrons by Gaussian basis sets. Both the 6-31G* and [PS]-41G* basis sets for O and Bi, respectively, can be found at CRYSTAL home page (http://www.crystal.unito.it/).

In order to study the stability of the α phase under pressure we have performed calculations not only for the α phase but also for the different structures (β, δ, A-type, HP, HPC, and R). The diagonalization of the Fock matrix was performed at adequate k-points grids in the reciprocal space, being the total number of k-points of 30, 18, 27, 13, 13, 12, and 30 for the α, β, δ, A-type, HP, HPC, and R phases, respectively. The use of different number of k-points is due to the fact that the primitive unit cells of the different phases contain different number of atoms. A proper choice of convergence tool parameters will result into achievement of the self consistent field cycle convergence. The FMIXING parameter, for example, permits to mix the Fock/Kohn-Sham matrix derivatives between the cycle n and the n-1 at a fixed percentage of cycle n-1. A 40 % of n-1 cycle mixing was used in our calculations. In the CRYSTAL program, five ITOL parameters control the accuracy of the calculation of the bielectronic Coulomb and exchange series, as well as the SCF convergence threshold on total energy and on density matrix. Selection is performed according to overlap-like criteria: when the overlap between two atomic orbitals is smaller than $10^{-ITOL}$, the corresponding integral is disregarded or evaluated in a less precise way. ITOL1 is the overlap threshold for Coulomb integrals, ITOL2 is the penetration threshold for Coulomb integrals, ITOL3 is the overlap threshold for HF exchange integrals, and ITOL4 & ITOL5 control the pseudo-overlap of the HF exchange series. Criteria for choosing the five tolerances are discussed in the CRYSTAL09 user's manual available at CRYSTAL home page (http://www.crystal.unito.it/). In our calculations ITOL1 to ITOL4 were set to $10^{-8}$ and ITOL5 to $10^{-14}$, assuring a convergence in total energy better than $10^{-7}$ Hartree in all cases.

In order to take into account the effect of pressure on the different phases of $Bi_2O_3$, we have optimized the geometrical parameters and the internal positions of all phases, at a number of fixed external pressures, ranging from -5 to 45 GPa. Then, the computed (E, P, V) values are used to minimize the enthalpy with respect to *V* at selected values of pressure in the range 0 to 45 GPa. In this respect, it must be noted that a phase is thermodynamically unstable with respect to another phase if the Gibbs free energy, G = E+PV-TS, of the latter is smaller than that of former at certain temperature and pressure. Since our calculations are performed at different pressures at T = 0 K, we only consider differences in enthalpy, H = E+PV, in order to check the possible phase transitions and therefore the stability of each phase.

## 4. Results and Discussion

### 4.1. High-pressure behavior of the structural parameters of the α phase

**Figure 2(a)** shows the ADXRD patterns of synthetic α-$Bi_2O_3$ with increasing pressure up to 22.2 GPa using Ar as PTM. The ADXRD pattern obtained for synthetic α-$Bi_2O_3$ at ambient pressure agrees well with the JCPDS data card No. 16-654. The measured lattice parameters at ambient conditions are: $a$ = 5.849(5) Å, $b$ = 8.164(8) Å, $c$ = 7.504(7) Å, and $\beta$ = 112.88(8)°, yielding a unit-cell volume $V_0$ = 330.1(6) Å$^3$. These values are in good agreement with those previously found in the literature **[15]**. Similar ADXRD measurements for the mineral bismite at ambient conditions yield values of $a$ = 5.848(6) Å, $b$ = 8.166(9) Å, $c$ = 7.509(8) Å and $\beta$ = 113.0(1)°, which results in a unit-cell volume $V_0$ = 330.1(7) Å$^3$. These values are in agreement with our *ab initio* calculations for the α phase, where we have found that $V_0$ is 6% underestimated in comparison with the experimental values.

ADXRD data can be assigned to α-$Bi_2O_3$ up to 20 GPa. In this pressure range, all diffraction peaks markedly shift to larger diffraction angles as pressure increases (see **Figure 2(a)**). At 6.6 GPa, Ar solidifies (*fcc* structure) and the peaks (111) and (200) related to this structure are detectable **[23,24]**. The Bragg peaks associated to Ar can be easily identified since its peaks have a different pressure evolution that those of the sample (Ar is much more compressible than $Bi_2O_3$). Using the peaks of solid Ar to verify the pressure measured through the rubies, it was observed that both scales differ by less than 1 GPa up to the maximum pressure reached in our experiment. As shown in **Figure 2(a)**, the x-ray diffraction peaks of the sample do not broaden considerably upon compression up to the pressure were amorphisation was detected (to be commented in the next section). This fact indicates that experimental conditions do not deviate considerably from quasi-hydrostaticity. This conclusion is also supported by the fact that the ruby fluorescence line widths were not affected much by compression up to 25 GPa.

The Rietveld refinement and the residuals at 0.1 GPa for the synthetic sample are shown in the **Figure 2(b)**. The residuals of the refinement are $R_p$ = 2.2%, $R_{wp}$ = 3.4%, and $\chi^2$ = 0.2. Similar residuals were obtained at all studied pressures. In the α phase, all atoms occupy 4*e* (x,y,z) Wyckoff sites; however, since O has a smaller x-ray scattering cross section than Bi, is difficult to accurately obtain the nine atomic positions corresponding to the three different oxygen atoms by Rietveld refinement of the ADXRD patterns at high pressures. Therefore, the original positions of the oxygen atoms were constrained at ambient pressure and only Bi fractional coordinates and unit-cell parameters were refined. In addition, since the site occupancy factors (SOF) and the atomic displacement factor (ADF) are correlated, and they are more sensitive to background subtraction than positional parameters, they were constrained to 1 and 0.5 Å$^2$, respectively, in order to reduce the number of free parameters used in the refinement [33]. **Table I** summarizes the atomic positions of Bi atoms obtained from refinement at 0.1 GPa which are in good agreement with those of the literature [32]. Taking into account the above considerations and the absence of relative changes of the intensities of the Bragg peaks with increasing pressure, we have found that the atomic coordinates of the two Bi atoms up to 20 GPa were similar to those at 0.1 GPa within experimental uncertainty. This result agrees with the weak pressure dependence of atomic parameters obtained from our theoretical calculations (not shown). In summary, we have neglected the pressure effect on the atomic positions [34], assuming those refined at 0.1 GPa, in order to extract the pressure evolution of the unit-cell parameters of the α phase up to 20 GPa.

**Figure 3** shows the pressure evolution of the unit-cell volume of α-$Bi_2O_3$ obtained from Rietveld refinements up to 20 GPa. The obtained *P-V* data are fitted using a third-order Birch-Murnaghan (BM) EOS to obtain the ambient pressure bulk modulus $B_0$ and its pressure derivative $B_0$' [35]. The unit-cell volume at zero pressure, the bulk modulus at zero pressure, and its pressure derivative are summarized in **Table II** and compared with the results obtained by our theoretical calculations. Also the implied value of the second derivative of the bulk modulus, $B_0$'', is given in **Table II** [36]. As can be observed, the bulk modulus of the synthetic sample ($B_0$ = 85.4(5) GPa) increase ~15% when Ar is substituted by MEW ($B_0$ = 98.1(1) GPa) as PTM. As it has been already observed in other materials, the use of different pressure media (which may produce different deviatoric stresses) affects the pressure dependence of the unit-cell volume, thus influencing the determination of values of $B_0$ [33,37-40]. This occurs basically because if deviatoric components are present in the Cauchy stress tensor, the sample under compression may suffer two simultaneous strains: a compression induced by hydrostatic pressure and an expansion caused by the Poisson effect. This fact may lead to an effective experimental compression smaller than when only hydrostatic pressure is present [41,42]. Note that differences in the unit-cell volume in the two experiments carried out in the synthetic sample become larger than error bars (which are smaller than the size of symbols in **Figure 3**) when pressure exceeds 10 GPa. On the other hand, it is noteworthy that the bulk modulus of mineral bismite ($B_0$ = 107.0(7) GPa) is ~25% larger than the bulk modulus of synthetic bismite under the same conditions (pressurized with Ar), thus indicating that the mineral sample is less compressible than the synthetic sample. Curiously, the value of $B_0$ for mineral bismite is close to that obtained in synthetic bismite through the shock wave technique ($B_0$ = 106 GPa) **[16]**. It must be stressed that, in general, these experimental values are in rather good agreement with our theoretical calculations (see solid lines in **Figure 3**) for the α phase ($B_0$ = 90.1(8) GPa).

It is important to note here that very different values for the pressure derivative of the bulk modulus are found in different experiments (see **Table II**). It is known that the bulk modulus and its pressure derivative are two parameters with a strong correlation **[43]**. Therefore, in order to properly compare the different reported bulk moduli, we have fit all available results to a second-order BM EOS with a fixed $B_0'$= 4 **[44]**. This is an approach that works well for comparing the compressibility data of many oxides in the pressure range covered by our experiments **[45,46]**. The difference in bulk compressibility for the three samples with fixed $B_0'$ follows the same trend as previously obtained when $B_0'$ is taken as free parameter in the EOS fit. The results are also in good agreement with shock-wave experiments when $B_0'$ is fixed to 4. Curiously, calculations slightly overestimate the bulk modulus ($B_0$ = 96.3(5) GPa) when $B_0'$ is fixed to 4. However, the observed difference in $B_0$ with respect to experimental values is typical of DFT calculations and consistent with their volume (bulk modulus) underestimation (overestimation) **[47]**.

In summary, bulk modulus of synthetic bismite is around 85.4 GPa, which is in good agreement with theoretical calculations (90.1 GPa) within both experimental and theoretical uncertainties. This value is near 25% smaller than that of natural bismite and that of synthetic bismite measured with shock wave techniques and 15% smaller than the bulk modulus of synthetic bismite measured with MEW. The much larger value of the bulk modulus for the mineral sample suggests that impurities present in the mineral sample affect the compressibility of $Bi_2O_3$. On the other hand, the deviation between 15% and 25% of the bulk moduli of synthetic samples studied under different PTM and with different techniques suggests that deviatoric stresses could influence the estimation of the compressibility of the material as observed in other compounds, like $BaWO_4$ **[33]** and $BaSO_4$ **[40]**. Regarding the influence of impurities in the crystal compressibility, we think that probably impurities could cause local defects in the crystal lattice which can locally reduce the crystal compressibility, leading to a reduction of the macroscopic compressibility (increase of the bulk modulus).

The bulk modulus of α-$Bi_2O_3$ can be compared with other related compounds. In particular, it can be compared to other metastable polymorphs of bismuth oxide. The bulk modulus of α-$Bi_2O_3$ is relatively higher than that of β-$Bi_2O_3$ (30 GPa) **[48]**, HP-$Bi_2O_3$ (32.8 GPa), and HPC-$Bi_2O_3$ (60.3 GPa) **[11,12]**. However, it should be noted that, for the HPC phase, a rather large value of $B_0'$ is reported **[12]**. Therefore, its bulk modulus cannot be directly compared with our experiments (with $B_0'$ < 4). In order to compare the bulk modulus of the HPC phase with our data, again we have fitted the data for the HPC phase of **Ref. 12** to a second-order EOS with $B_0$' fixed to 4 (see **Table II**). In that way, we have found that the HPC phase is less compressible ($B_0$ = 99.3(4) GPa) than the α phase. This result is consistent with the fact that the HPC phase has a more compact and denser volume and that the HPC phase is a stable structure at high pressures (even a possible post α phase) as will be commented in the next section. Finally, the bulk modulus of α-$Bi_2O_3$ can be compared to that of other V-group sesquioxides. The bulk modulus of α-$Bi_2O_3$ is significantly larger than that of arsenolite (cubic $As_2O_3$ - 18 GPa) **[18]** and than that of senarmontite (cubic $Sb_2O_3$ - 20 GPa) **[20]**, both being molecular crystals. Unfortunately, comparison with claudetite (monoclinic $As_2O_3$) and valentinite (orthorhombic $Sb_2O_3$) is not possible because the EOS of both compounds has not been reported to our knowledge.

X-ray data analysis also allows us to estimate the pressure dependence of the lattice parameters (*a*,*b*,*c*) and the *β* angle (see **Figure 4**). Axial compressibilities at zero pressure have been estimated from a fit of experimental data to a modified Murnaghan EOS (see **Table II**) **[49]**. The compressibility of the *b* axis in α-$Bi_2O_3$ is higher than those of the *a* and *c* axes in the three experimental sets. This behavior is consistent with our theoretical calculations (see solid lines in **Figure 4(a)**). On the other hand, the *a*, *b* and *c* axial compressibility of synthetic α-$Bi_2O_3$ is ~50%, ~15% and ~29% higher than the ones for mineral sample, respectively, under the same hydrostatic conditions. Finally, the results presented for the synthetic sample in **Table II** also show that the use of MEW as PTM compared to Ar produce a decrease in axis compressibility, mainly in the *a* and *c* axis. Curiously, the anisotropic compressibility of the different axes is comparable with that observed in $PbCrO_4$ which also has a monoclinic structure **[50]**.

An interesting issue related with axial compressibilities of α-$Bi_2O_3$ is that at 20 GPa, *b* and *c* lattice parameters become nearly equal in value (**Figure 4(a)**). Noteworthy, this value is similar to the value of the *a* and *b* axes in hexagonal HPC-$Bi_2O_3$ (7.092 Å **[12]**). Furthermore, the value of the *a* axis of α-$Bi_2O_3$ also takes approximately the same value at 20 GPa than the *c* axis in HPC-$Bi_2O_3$ (5.856 Å **[12]**). These facts can be an indication that pressure gradually converts the monoclinic α-$Bi_2O_3$ structure into a pseudo-hexagonal structure with some structural similarities to the hexagonal HPC-$Bi_2O_3$ around 20 GPa; however, the transformation to the HPC phase would require that *β* angle tend to 90º around 20 GPa (which is not the case). The tendency of the lattice parameters of the α phase towards those of the HPC phase is consistent with our theoretical calculations that show a higher stability of the HPC phase than the α phase at high pressures as will be commented in the next section. On the other hand, the lack of tendency of the *β* angle towards 90º could be a signature of the inability of the α phase to undergo the transition to the HPC phase at room temperature.

From analysis of ADXRD data it was also possible to obtain information on the compressibility of interatomic distances. **Figure 5** shows the pressure dependence of the experimental cation-anion and cation-cation interatomic distances for the synthetic sample pressurized with Ar. Similar results were found for the pressure dependence of the interatomic distances in the other experiments (not shown). **Table III** summarizes the compressibility of the interatomic distances at zero pressure in the different experiments and compare them with those obtained from our theoretical calculations. The results indicate that the PTM type (Ar or MEW) did not influence significantly the compressibility of the Bi-O binding distances in the synthetic sample. However, the comparison of the synthetic and the mineral sample (both pressurized with Ar) allows us to observe that all interatomic distances have lower pressure coefficients in the mineral sample. It is also possible to observe that, on average, the bonds of $BiO_5$ polyhedra are less compressible than those of $BiO_6$ polyhedra. Furthermore, the separation between the shortest and the largest bond distances in both $BiO_5$ and $BiO_6$ units increase with pressure, thus evidencing that these units become more irregular under compression.

*4.2. Amorphisation of the α phase under pressure*

In order to get further insight into the possible pressure-induced transformations of α-$Bi_2O_3$, we have performed total-energy *ab initio* calculations of several phases of $Bi_2O_3$ (α, β, δ, A-type, HP, HPC, and R) found at different pressures and temperatures **[3,11,12,14]**. The aim was to check the stability of the α phase with respect to other phases which could be candidates to high-pressure phases of bismite. **Figure 6** shows the pressure dependence of the enthalpy difference (relative to the α phase) for the HP and HPC phases, which are the only ones that are competitive with the α phase at high pressures. The negative values of the theoretical enthalpy difference for the HP-$Bi_2O_3$ and HPC-$Bi_2O_3$ phases with respect to the α phase above 5.5 GPa indicate that the polymorphs HP-$Bi_2O_3$ and HPC-$Bi_2O_3$ are energetically more stable than the α-$Bi_2O_3$ above 5.5 GPa, in good agreement with the results of Ghedia *et al.* **[11]** and Locherer *et al.* **[12]**. Furthermore, these authors showed experimentally that the HP phase was unstable with respect to the HPC phase at ambient temperature above 3 GPa **[12]**. This result is in good agreement with our calculations and would suggest the possibility of a phase transition from the α phase directly to the HPC phase above 5.5 GPa.

Upon compression of the synthetic sample of α-$Bi_2O_3$ with Ar above 20 GPa, the Bragg peaks lose their shapes at 22.2 GPa, and only broad bands corresponding to diffuse x-ray scattering are observed (see **Figure 2(a)**). These bands suggest either the amorphisation of the material or the formation of a glass **[51,52]** instead of the transformation to the HPC phase; although the lattice parameters of the α phase at 20 GPa are close to the values of lattice parameters of the HPC phase at 20 GPa, as already noted in the previous section. Our results are in agreement with the amorphisation of α-Bi$_2$O$_3$ above 20 GPa suggested by Chouinard *et al.* from Raman scattering measurements [15]. Furthermore, we also observed the amorphisation in mineral α-Bi$_2$O$_3$ using Ar and in synthetic α-Bi$_2$O$_3$ using MEW at 25 GPa and at 15 GPa, respectively. In this last sample, pressure was increased up to 45 GPa but no major change of the amorphous phase was detected except for a small shift of the bands to higher angles (smaller interplanar distances, see inset of **Figure 7**). This shift is a consequence of the decrease of bond distances under compression.

Our three experiments evidence, on one hand, that amorphisation of α-Bi$_2$O$_3$ takes place in mineral bismite at a higher pressure than in the synthetic pure sample. A similar behavior was earlier observed in zircon [53]. This observation suggests that impurities present in the mineral oxides affect the amorphisation kinetics of α-Bi$_2$O$_3$ and by analogy perhaps it could affect the amorphisation process in other sexquioxides. To further explore, whether there is a systematic effect of impurities on the kinetics of phase transitions of Bi$_2$O$_3$ (and its compressibility), additional high-pressure x-ray diffraction measurements on Bi$_2$O$_3$ samples with well-known compositions are clearly needed. On the other hand, they evidence that deviatoric stresses reduce the amorphisation pressure in α-Bi$_2$O$_3$ since amorphisation takes place at lower pressures in a less hydrostatic environment. Again, a similar behavior was also observed in BaWO$_4$ [33] and BaSO$_4$ [39].

In order to obtain more information on the amorphous phase we have plotted in **Figure 7** the diffractogram of synthetic α-Bi$_2$O$_3$ pressurized with Ar at 22.2 GPa in the amorphous phase but with the 2θ coordinate converted into interplanar distances [52]. Narrow peaks corresponding to solid Ar (at small distances) and three main broad bands (at 2.27, 2.81, and 3.22 Å) can be observed in the diffraction pattern which corresponds to possible interatomic distances in the amorphous phase. **Table IV** summarizes the main interatomic distances at 22.2 GPa in the amorphous material compared to those of the HPC phase [12] and the α phase (this work) at a similar pressure. According to data for the HPC phase, the smallest Bi-O distances are in the range of 2.0 to 2.6 Å (average 2.36 Å), the smallest O-O distances are in the range of 2.7 to 3.3 Å (with eight out of fourteen distances between 2.7 and 2.9 Å), and the smallest Bi-Bi distances are below 3.29 Å. On the other hand, in the α phase there is a much larger dispersion of interatomic distances than in the HPC phase, being the smallest Bi-O distances in the range of 1.9 to 2.7 Å (average 2.24 Å), the smallest O-O distances are in the range of 2.6 to 3.7 Å (average 3.09 Å), and the smallest Bi-Bi distances are in the range of 3.26 to 4.16 Å (average 3.64 Å). These data reveal that the average Bi-O interatomic distances in the HPC phase are slightly larger than those of the α phase at 20 GPa what is consistent with the larger Bi coordination of the HPC phase (average 6.5) with respect to the α phase (5.5).

On the basis of the above comparison of interatomic distances, we suggest that the interatomic distances of the broad peaks in the diffraction pattern of the amorphous phase at 22.2 GPa likely correspond to those of the smallest interatomic distances of the HPC phase; i.e., the amorphous phase seems to be a poorly crystallized HPC phase. The main feature for this assignment is the narrow and intense band at 2.82 Å which can be clearly assigned to the O-O distances in the HPC phase because many O-O distances in this structure lay in a very narrow range between 2.7 and 2.9 Å near 20 GPa. Furthermore, the third broad band whose maximum is around 3.22 Å likely corresponds to the smallest Bi-Bi distance in the HPC phase (3.29 Å at 20 GPa). Note that larger values of Bi-Bi distances would be expected in the α phase (around 3.41 Å on average and beyond). Finally, the first broad band in the amorphous phase which has a maximum at 2.27 Å and a plateau for slightly higher energies could be attributed to Bi-O distances in the HPC phase which lay between 2.0 and 2.6 Å (average of 2.36 Å). Again we must note that a more symmetric band with a maximum at 2.24 Å would be expected for Bi-O distances in the α phase at 20 GPa. Finally, the lack of peaks in the XRD pattern above 3.5 Å can be considered as an indication that this phase is amorphous since the constructive interference disappears for high distances in the amorphous phase because of the lack of long range order.

Pressure-induced amorphisation (PIA) occurs at relatively low temperatures in a number of compounds that were predicted to undergo a phase transition to a crystalline phase **[54-57]**. There is a long-standing controversy about whether PIA is of a mechanical or thermodynamical nature and its relation to the two possible mechanisms of melting at high temperatures **[58-61]**. In this respect, PIA was originally explained as a metastable melting **[62]** but later as a mechanical melting driven by elastic or lattice instabilities **[63-65]**. In general, several mechanisms for PIA have been proposed where defects and non-hydrostatic stresses usually play an important role **[54-57,59-70]**, and where the main models consider the amorphous phase as a consequence of a frustrated transition from a parent crystalline phase to another crystalline phase **[55]**. For instance, according to theoretical predictions, trigonal AlPO$_4$ has a phase transition to the orthorhombic *Cmcm* structure above 10 GPa. However, either crystalline-to-crystalline or crystalline-to-amorphous transitions have been observed in this compound under different hydrostatic conditions and at different temperatures **[71-73]**. On the basis of the above results, the crystal-to-amorphous phase transition observed in α-Bi$_2$O$_3$ results in an increase of the Bi coordination from 5.5 to 6.5 so it seems to be similar to that reported in arsenolite (α-As$_2$O$_3$) **[18]**, where PIA was suggested to be related to an increase in the coordination number of As, as suggested by the increase in the average As-O bond length after amorphisation. We note that in order to better characterize the local atomic structure of the amorphous phase HP x-ray absorption spectroscopy and high-energy x-ray diffraction measurements are advisable.

PIA in α-Bi$_2$O$_3$ lead to the observation of an interesting phenomenon: the samples changed their color from light yellow to dark red, almost black. This phenomenon has already been observed both in Bi$_2$O$_3$ **[15]** as in As$_2$O$_3$ glass **[18]** and can be indicative of a collapse of the bandgap which can lead to a major change in its electronic properties. One possible explanation of the band-gap collapse is that it could be caused by the high distortion of the BiO$_5$ and BiO$_6$ polyhedra induced after PIA. This fact will lead to changes in the electronic density around Bi, which should be directly reflected in the electronic structure of Bi$_2$O$_3$, as observed in other oxides **[50,74].** However, an accurate determination of the causes of this phenomenon is beyond the scope of this work.

PIA in α-Bi$_2$O$_3$ is likely related to the impossibility to undergo a crystalline-to-crystalline phase transition to the HPC phase, as it occurs in other compounds **[55,75]**. The difficulty of α-Bi$_2$O$_3$ to transform into HPC-Bi$_2$O$_3$ at high pressures and room temperature is likely due to the existence of a high energy barrier between both structures that cannot be overcome only by applying pressure. Note that α-Bi$_2$O$_3$ needs to be pressurized to 6 GPa and 900ºC to undergo a phase transition to HPC-Bi$_2$O$_3$ **[11]**. In this scenario, the amorphous phase is a metastable phase, which is energetically more stable and kinetically advantageous when compared to the high-pressure HPC polymorph **[55]**.

In order to get a better insight into the PIA process, we have calculated the elastic constants of α-Bi$_2$O$_3$ as a function of pressure up to 25 GPa. Our results on the calculated elastic constants, which will be published elsewhere, indicate that the crystalline structure of α-Bi$_2$O$_3$ becomes mechanically unstable above 19 GPa as a consequence of the violation of the generalized Born stability criteria **[76]**. Therefore, in our opinion, PIA in α-Bi$_2$O$_3$ above 20 GPa takes place because: i) the α phase is unstable with respect to the HPC phase above 5 GPa; ii) Bi atoms cannot reach the atomic positions in the HPC structure above 20 GPa despite Bi-O, O-O, and Bi-Bi distances are similar to those present in the crystalline HPC phase at the same pressure; and iii) the α phase becomes mechanically unstable above 19 GPa. This sequence of phenomena causes the final collapse of the structure to yield the amorphous phase above 20 GPa which seems to be a poorly crystallized HPC phase. In a forthcoming paper we will discuss the mechanism of PIA and will show that an increase of temperature at pressures above 20 GPa result in the crystallization of the HPC phase from the original α phase **[76]**.

Considering the hypothesis of an impeded transition from the α phase to the HPC phase as the initial cause of PIA, an interesting question that arises is why HP-Bi$_2$O$_3$ transits to HPC-Bi$_2$O$_3$ at relative low pressure (~3 GPa) and ambient temperature **[12]** while α-Bi$_2$O$_3$ cannot undergo a phase transition to HPC-Bi$_2$O$_3$ beyond 5.5 GPa, but it can undergo a phase transition to HPC-$Bi_2O_3$ at 6 GPa and 900ºC **[11,12]**. The answer to this question can be directly related to the crystalline structures of these polymorphs. **Figure 8** may help one to understand the phase transition mechanisms for the different polymorphs of $Bi_2O_3$. In **Figure 8**, four connected polyhedra are shown: one $BiO_6$ unit for each structure, three $BiO_5$ units for α and HP structures and three $BiO_7$ units for HPC structure. The α-HPC transition, which occurs at ~6 GPa and 900ºC, seems to be a consequence of the torsion of $BiO_5$ units with respect to the $BiO_6$ unit in a continuous way that leads from the a phase to the HPC phase via the intermediate HP phase **[11]**. In this respect, the transition HP-HPC is a result of torsion of the $BiO_6$ polyhedron, thus inducing the formation of a plane mirror and, consequently, the $BiO_5$ polyhedra undergo a tilt and approach each other. Each Bi of these units bind with two oxygens of a neighbor $BiO_5$ polyhedron, thus forming $BiO_7$ units in the HPC phase **[12]**. Therefore, the α-HPC transition occurs through a sequence of α-HP and HP-HPC transitions. In this way, at low temperatures the same kinetic reasons that impede the α-HP transition impede the α-HPC transition. The polyhedral torsions and atomic bonds which are needed to turn the α phase into the HP and HPC phases seem to be too complex, as indicated by the inability of the *β* angle of the α phase to reach 90º (see **Figure 4(b)**). Thus, it is reasonable to think that the system does not have enough energy to overcome the kinetic barriers at ambient temperature. However, the increase of temperature to 900ºC above 6 GPa allows the α-HPC transition **[11,12]**.

Finally, we must note that after increasing pressure to 22.2 GPa and 25 GPa in the synthetic and mineral sample, respectively, we decreased pressure slowly down to ambient pressure and observed the reversibility of the PIA process in both samples (see top of **Figure 2(a)**); however, in the sample pressurized with MEW up to 45 GPa, after a non-gradual pressure release, the amorphous state was quenched to ambient pressure. These results compare to those obtained with synthetic samples and MEW by Chouinard *et al.* **[15]**. They found an irreversibility of the crystalline-to-amorphous transition above 20 GPa upon decompression but recovered the crystallinity at ambient pressure after thermal annealing. These results altogether suggest that the reversibility of PIA is influenced by deviatoric stresses, which are known to strongly influence structural changes **[71]**. Probably, the PIA process is not reversed upon decompression only when non-hydrostatic stresses frustrate the recrystallization of the thermodynamically stable phase through the enhancement of kinetic barriers which are overcome by applying temperature on the annealing process **[77]**.

According to studies performed in other oxides **[78,79]**, the recovering of amorphised structures can be related to the presence of non-deformed units of the initial phase. In this sense, the presence of undeformed units after PIA ($BiO_6$ units in $Bi_2O_3$), added to the fact that the pressure was applied by a quasi-hydrostatic PTM and released slowly, may be one of the factors responsible for the recovery of the initial crystalline structure. In the case of measurements where MEW was used as PTM, in addition to being a less hydrostatic media, the sample was quenched rapidly, disabling the recovery of the crystalline structure.

## 5. Conclusions

We report a room-temperature ADXRD study of synthetic and mineral bismite ($\alpha$-$Bi_2O_3$) at high pressures. The experimental equation of state of the studied samples is in good agreement with that obtained from *ab initio* calculations and recent experiments with shock waves. It was observed that the bulk modulus of the synthetic sample increases ~15% when Ar was substituted by a less hydrostatic pressure-transmitting medium. Besides, there is an increase of ~25% in the bulk modulus in the mineral sample when compared to the synthetic under the same pressure conditions. These results suggest that both the impurities of the mineral sample and a less hydrostatic pressure-transmitting medium reduce the compressibility of $\alpha$-$Bi_2O_3$.

The amorphisation of bismite occurs in the range between 15 and 25 GPa which depends on the quality of the sample and the pressure-transmitting medium. The amorphisation process seems to be reversible using Ar and not reversible using methanol-ethanol-water. Theoretical calculations indicate that the crystalline structure of $\alpha$-$Bi_2O_3$ becomes unstable against HPC-$Bi_2O_3$ above 5.5 GPa and that the $\alpha$ phase becomes mechanically unstable above 19 GPa as a consequence of the violation of the generalized Born stability criteria. Therefore, the pressure-induced amorphisation process of $\alpha$-$Bi_2O_3$ at room temperature seems to be a consequence of the inability of the $\alpha$ phase to undergo a phase transition to another crystalline phase, likely the HPC phase **[12]**. Furthermore, the amorphous phase seems to be a poorly crystallized HPC phase. New studies of $\alpha$-$Bi_2O_3$ above 20 GPa and at high temperatures are needed to verify if the crystallization of the HPC phase can be attained directly from the $\alpha$ phase.

**Acknowledgements**

Financial support from the Spanish Consolider Ingenio 2010 Program (MALTA Project No. CSD2007-00045) is acknowledged. This work was also supported by Brazilian Conselho Nacional de Desenvolvimento Científico e Tecnológico (CNPq) under project 201050/2012-9, Spanish MICINN under projects MAT2010-21270-C04-01/03/04, Spanish MINECO under project CTQ2012-36253-C03-02, and from Vicerrectorado de Investigación de la Universitat Politècnica de València under projects UPV2011-0914 PAID-05-11 and UPV2011-0966 PAID-06-11. Supercomputer time has been provided by the Red Española de Supercomputación (RES) and the MALTA cluster.

Table I. Atomic coordinates of synthetic α-Bi$_2$O$_3$ obtained from Rietveld refinement of powder diffraction at 0.1 GPa. Oxygen positions were taken from literature data at ambient pressure **[32]** and were not refined.

| Atom | Site | x | y | z |
| --- | --- | --- | --- | --- |
| Bi I | 4*e* | 0.5294(6) | 0.1963(2) | 0.3597(4) |
| Bi II | 4*e* | 0.0365(1) | 0.0550(1) | 0.7772(8) |
| O I | 4*e* | 0.7770 | 0.3040 | 0.7070 |
| O II | 4*e* | 0.2350 | 0.0480 | 0.1270 |
| O III | 4*e* | 0.2690 | 0.0280 | 0.5110 |

Table II. EOS parameters and axial compressibility ( —— ) at ambient pressure of synthetic and natural α-Bi$_2$O$_3$ obtained under different pressure-transmitting media (PTM). The variation — was obtained using the Murnaghan equation of state ——, were and are the bulk modulus and its pressure derivative of the $x$-axis ($x=a, b, c$) at atmospheric pressure.

| Sample (PTM) | $V_0$ (Å$^3$) | $B_0$ (GPa) |  | (GPa$^{-1}$) | $\kappa_a$ (10$^{-3}$ GPa$^{-1}$) | $\kappa_b$ (10$^{-3}$ GPa$^{-1}$) | $\kappa_c$ (10$^{-3}$ GPa$^{-1}$) |
|---|---|---|---|---|---|---|---|
| Synthetic (Ar) | 329(1) | 85.4(5) | 2.6(5) | -0.052 | 2.07(1) | 6.64(1) | 4.41(1) |
|  | 329(1) | 71.7(3) | 4.0 (fixed) | -0.054 |  |  |  |
| Synthetic (MEW) | 330(1) | 98.1(1) | 1.7(1) | -0.070 | 1.15(1) | 6.21(2) | 3.20(1) |
|  | 330(1) | 79.2(3) | 4.0 (fixed) | -0.049 |  |  |  |
| Mineral (Ar) | 330(1) | 107.0(7) | 1.6(5) | -0.068 | 1.02(1) | 5.64(2) | 3.15(2) |
|  | 330(2) | 86.4(6) | 4.0 (fixed) | -0.045 |  |  |  |
| Theoretical | 310.2(1) | 90.1(8) | 4.8(1) | -0.059 | 1.53(1) | 7.84(2) | 2.50(1) |
|  | 309.7(1) | 96.3(5) | 4.0 (fixed) | -0.040 |  |  |  |
| HP-Bi$_2$O$_3$ (He)[a] | 328(2) | 32.8(26) | 6.2(37) | -0.330 |  |  |  |
|  | 328(1) | 34.5(2) | 4.0 (fixed) | -0.110 |  |  |  |
| HPC-Bi$_2$O$_3$ (He)[a] | 308(1) | 60.3(30) | 8.1(3) | -0.410 |  |  |  |
|  | 302(1) | 99.3(4) | 4.0 (fixed) | -0.039 |  |  |  |
| α-Bi$_2$O$_3$ (shock wave)[b] |  | 106 | 1.28 | -0.080 |  |  |  |
|  |  | 82 | 4.0 | -0.047 |  |  |  |

[a] Ref. 12
[b] Ref. 16

Table III. Pressure coefficients for the cation-anion and cation-cation interatomic distances in α-Bi$_2$O$_3$ for our three different experiments and theoretical calculations.

|  | (10$^{-3}$ Å/GPa) | | | |
| --- | --- | --- | --- | --- |
|  | Synthetic (Ar) | Synthetic (MEW) | Mineral (Ar) | Theoretical |
| Bi I-O I | -8.7 | -8.6 | -8.2 | -3.6 |
| Bi I-O II | -5.5 | -5.5 | -4.8 | -4.4 |
| Bi I-O III | -3.6 | -3.7 | -2.6 | -4.6 |
| Bi I-O I' | -6.6 | -6.6 | -5.6 | -5.1 |
| Bi I-O III' | -6.6 | -6.8 | -5.9 | -6.0 |
| Bi II-O I | -3.9 | -3.9 | -3.0 | -3.9 |
| Bi II-O II | -10.1 | -9.9 | -9.5 | -5.0 |
| Bi II-O III | -4.2 | -4.3 | -3.5 | -3.7 |
| Bi II-O I' | -5.3 | -5.7 | -4.8 | -6.6 |
| Bi II-O II' | -10.6 | -10.4 | -9.8 | -7.9 |
| Bi II-O III' | -5.5 | -5.8 | -4.5 | -4.9 |
| Bi I-Bi II | -9.5 | -9.2 | -8.2 | -8.2 |

Table I. Interatomic distances (in Å) obtained in the high-pressure amorphous phase of α-$Bi_2O_3$ at 22.2 GPa (from Figure 7) compared to those of HPC-$Bi_2O_3$ estimated at 20 GPa from Ref. 12 and with those of α-$Bi_2O_3$ at 20 GPa. Values in parenthesis indicate the number of equal (degenerate) interatomic distances.

| | | | |
|---|---|---|---|
| **Amorphous $Bi_2O_3$ (22.2 GPa)** | | | |
| Peak 1 | Peak 2 | | Peak 3 |
| 2.27 | 2.81 | | 3.22 |
| **HPC-$Bi_2O_3$ (20 GPa)** | | | |
| Bi-O ($BiO_6$) | Bi-O ($BiO_7$) | O-O | Bi-Bi |
| 2.1285 (3) | 2.0302 (2) | 2.6951 (2) | 3.2873 (3) |
| 2.5640 (3) | 2.3273 (1) | 2.7022 (2) | 3.3902 (2) |
| | 2.5127 (2) | 2.7607 (2) | 3.5589 (6) |
| | 2.5817 (2) | 2.8780 (2) | 3.6252 (4) |
| | | 3.0885 (2) | 3.7022 (2) |
| | | 3.3177 (4) | 3.7639 (1) |
| | | 3.8606 (2) | |
| **α-$Bi_2O_3$ (20GPa)** | | | |
| Bi-O ($BiO_6$) | Bi-O ($BiO_5$) | O-O | Bi-Bi |
| 2.0089 (1) | 1.8980 (1) | 2.6472 (1) | 3.2654 (1) |
| 2.0558 (1) | 2.0591 (1) | 2.6514 (1) | 3.3698 (1) |
| 2.2043 (1) | 2.1325 (1) | 2.7375 (1) | 3.4158 (1) |
| 2.3148 (1) | 2.4062 (1) | 2.7747 (1) | 3.4207 (1) |
| 2.3638 (1) | 2.5005 (1) | 2.7750 (1) | 3.4431 (1) |
| 2.6792 (1) | | 2.9802 (1) | 3.4536 (1) |
| | | 2.9996 (1) | 3.6308 (2) |
| | | 3.0901 (1) | 3.6931 (2) |
| | | 3.1108 (1) | 3.7999 (1) |
| | | 3.1303 (1) | 3.8298 (1) |
| | | 3.3626 (1) | 3.8334 (1) |
| | | 3.6373 (2) | 3.9010 (1) |
| | | 3.6645 (1) | 4.1547 (1) |

**Figure captions**

Figure 1. (color online) Crystalline structure of monoclinic $\alpha$-$Bi_2O_3$ at ambient pressure view towards plane (0-10). Grey balls represent Bi atoms, while red balls represent O atoms. The structure has one Bi with coordination five (Bi-I – green polyhedra) and another with coordination six (Bi-II – blue polyhedra).

Figure 2. (a) Room-temperature powder x-ray diffraction patterns of synthetic $\alpha$-$Bi_2O_3$ measured at different pressures (spectra are shifted vertically for increasing pressures). The top pattern corresponds to a pattern collected in a recovered (r) sample at 0.1 GPa after decompression from 22.2 GPa thus showing the reversibility of the pressurization process. Asterisks in patterns above 6.6 GPa are related to peaks of solid Ar. (b) Powder XRD pattern measured at 0.1 GPa shows the Rietveld refined spectrum (dotted line) and residues (lower line).

Figure 3. Unit-cell volume vs. pressure for $\alpha$-$Bi_2O_3$. Symbols and solid line represent experimental and theoretical data obtained by *ab initio* calculation, respectively. Dashed, dotted, and dash-dotted lines are the result of the 3$^{rd}$ order Birch-Murnaghan EOS fit to experimental data. The error bars are comparable with symbols sizes.

Figure 4. Pressure evolution of the (a) lattice parameters and (b) $\beta$ angle. Symbols and solid line represent experimental and theoretical data, respectively. Dashed, dotted, and dash-dotted lines correspond to fits of a Murnaghan EOS for the lattice parameters and to quadratic fits for the $\beta$ angle. The error bars are comparable with symbols sizes.

Figure 5. Cation-anion and cation-cation distances obtained from synthetic $\alpha$-$Bi_2O_3$ (Ar). The index of each atom is represented in Figure 1.

Figure 6. Theoretical calculation of the enthalpy difference as a function of pressure for the $\alpha$, HP, and HPC phases of $Bi_2O_3$. Enthalpy of phase $\alpha$ is taken as the reference.

Figure 7. Difractogram of amorphous synthetic $Bi_2O_3$ taken at 22.2 GPa using Ar as PTM as a function of the interplanar distance. Inset: Diffractogram of amorphous synthetic $Bi_2O_3$ taken at 26.2 and 45.3 GPa using MEW as PTM as a function of the interplanar distances. The diffractogram present an intense and narrow peak corresponding to the tungsten gasket.

Figure 8. (color online) Sequence of pressure and temperature induced phase transition in $Bi_2O_3$. Bi atoms are the gray bigger balls, while O atoms are red smaller balls. RT corresponds to room temperature.

**Figure 1**

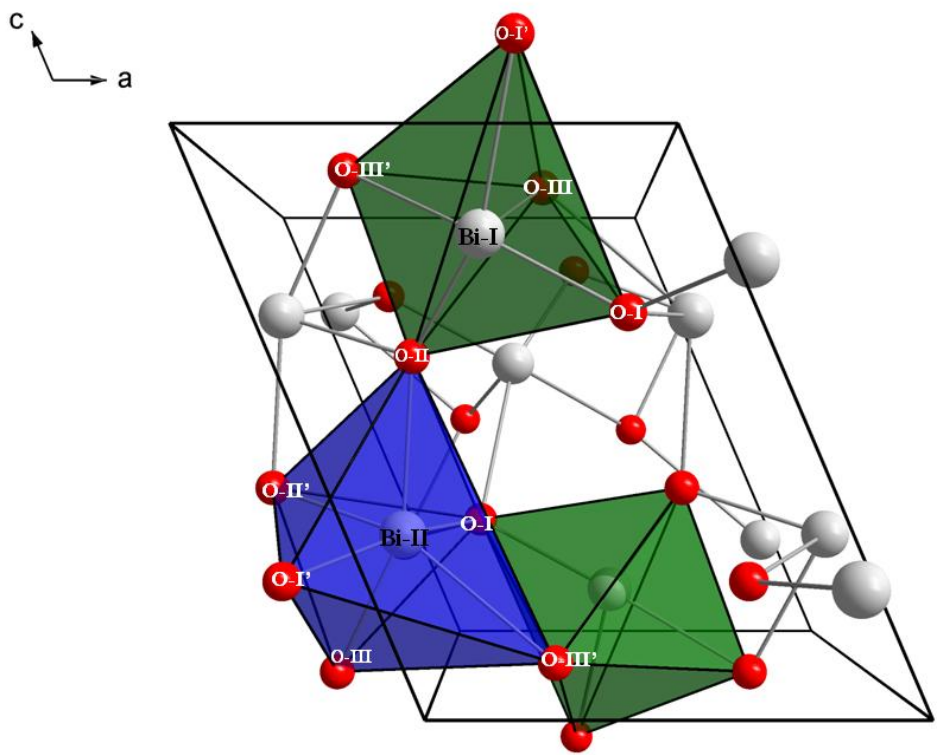

**Figure 2**

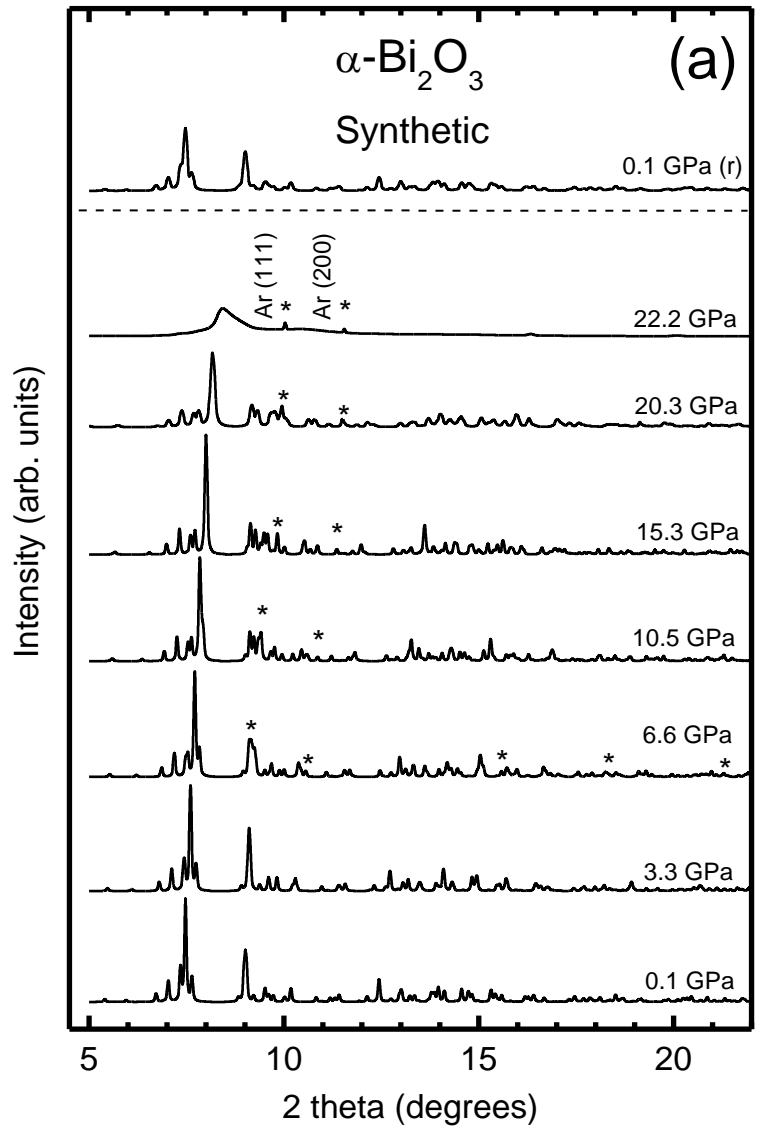

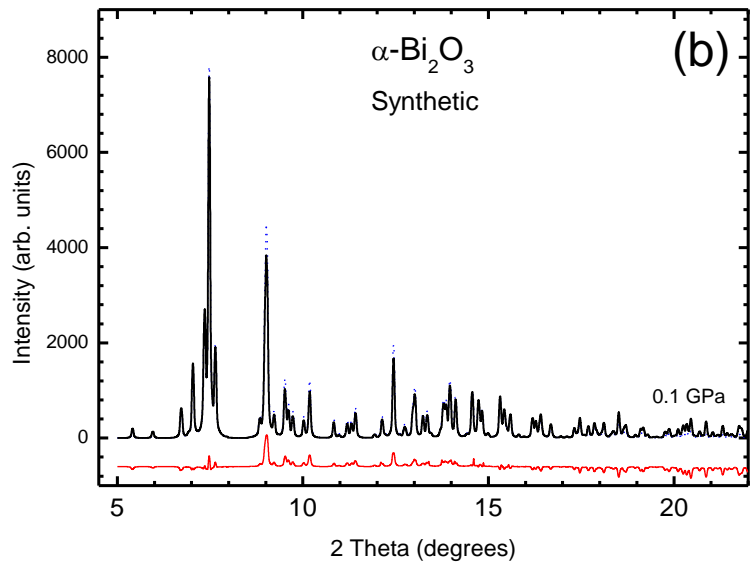

**Figure 3**

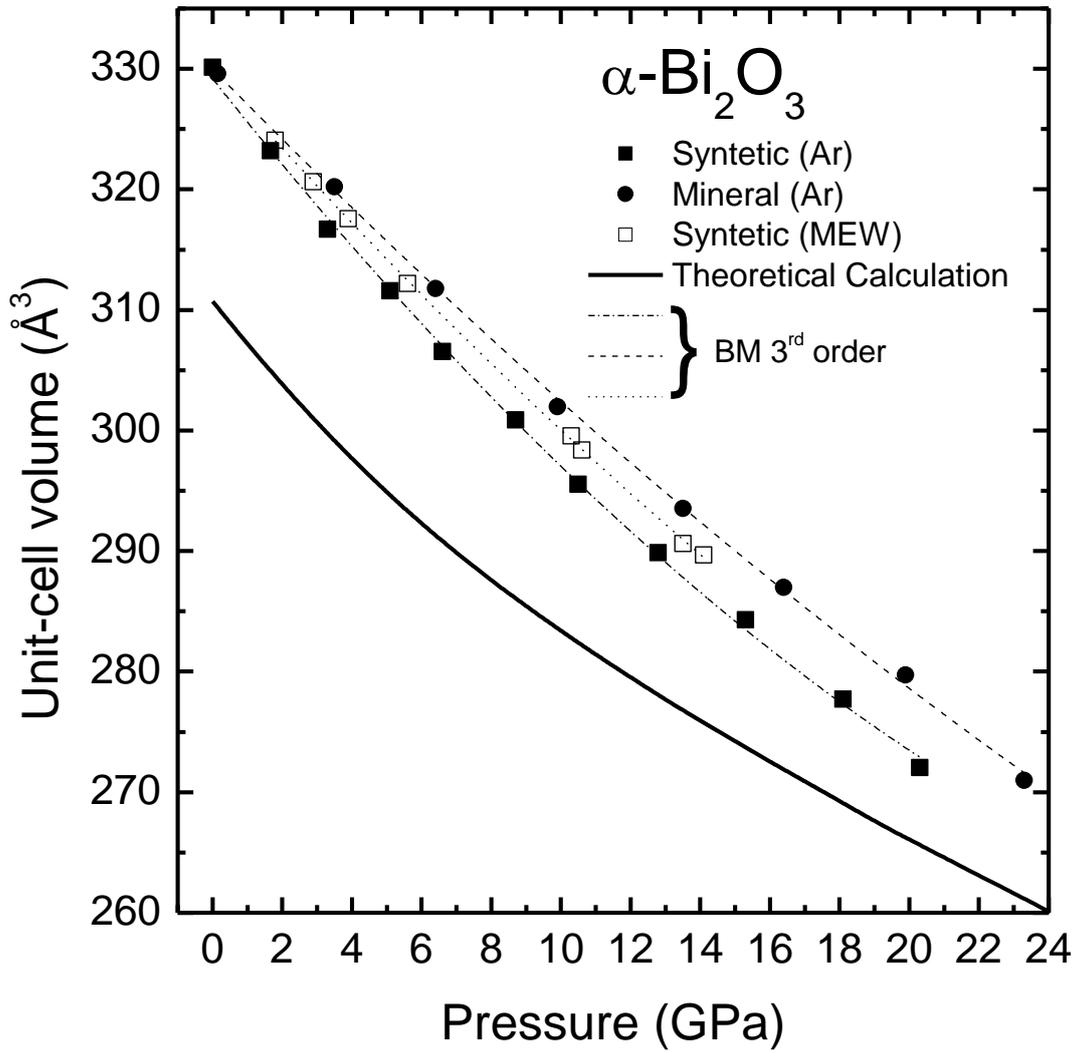

**Figure 4**

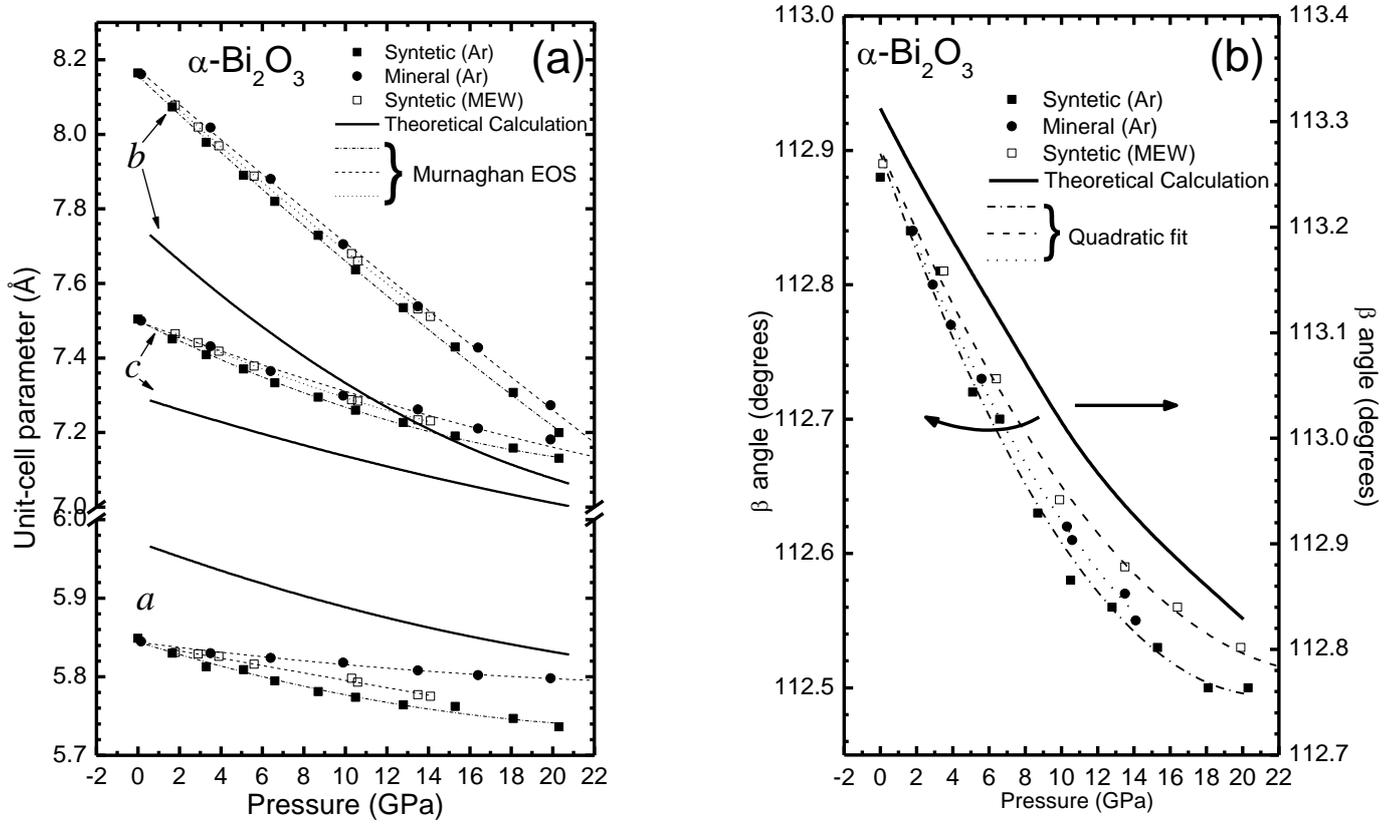

**Figure 5**

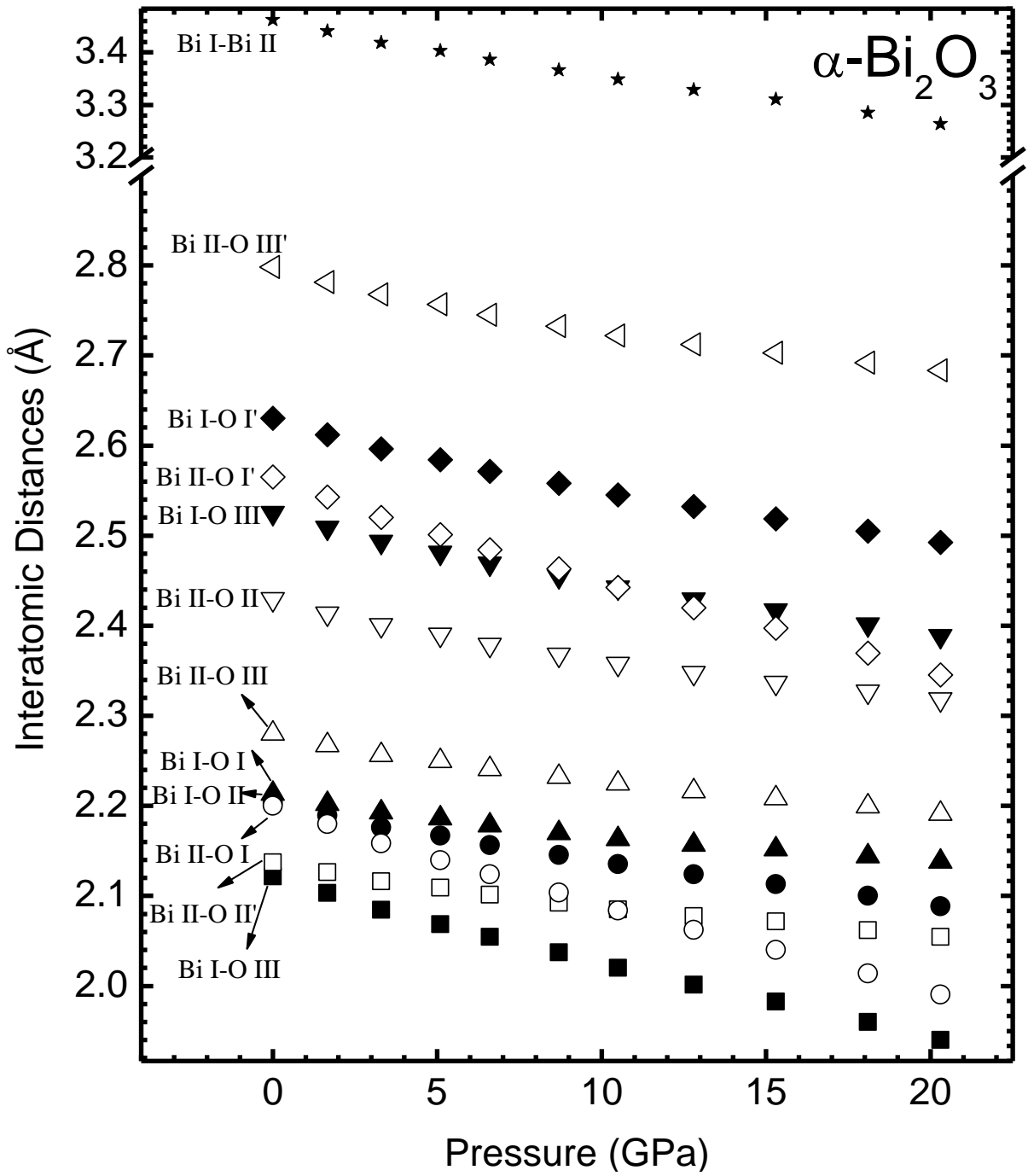

**Figure 6**

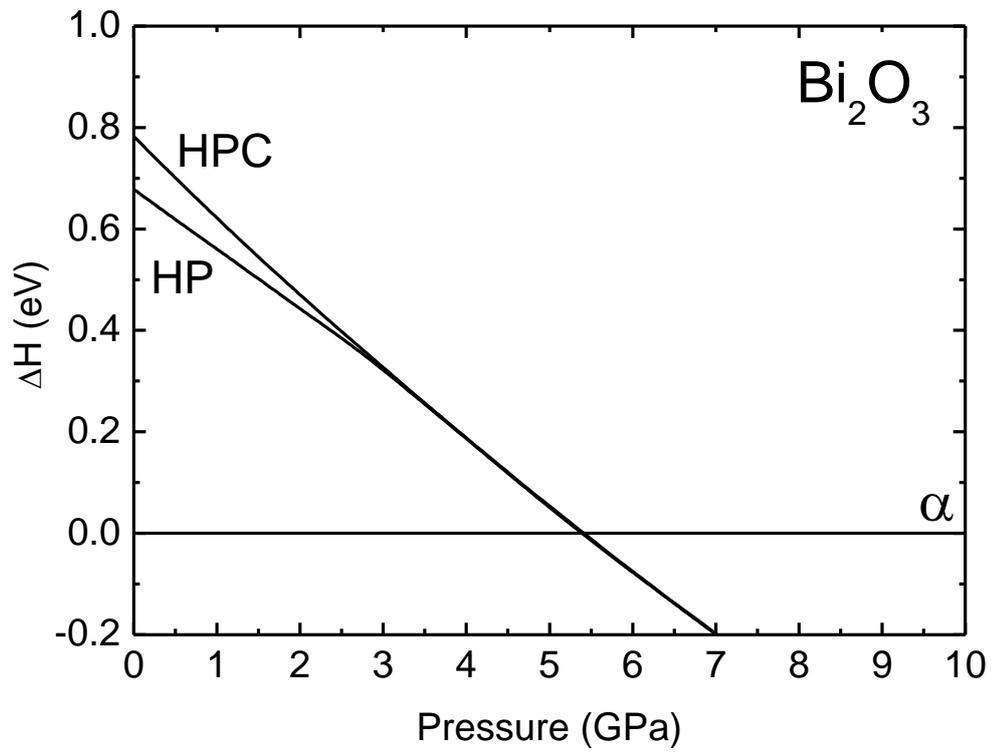

**Figure 7**

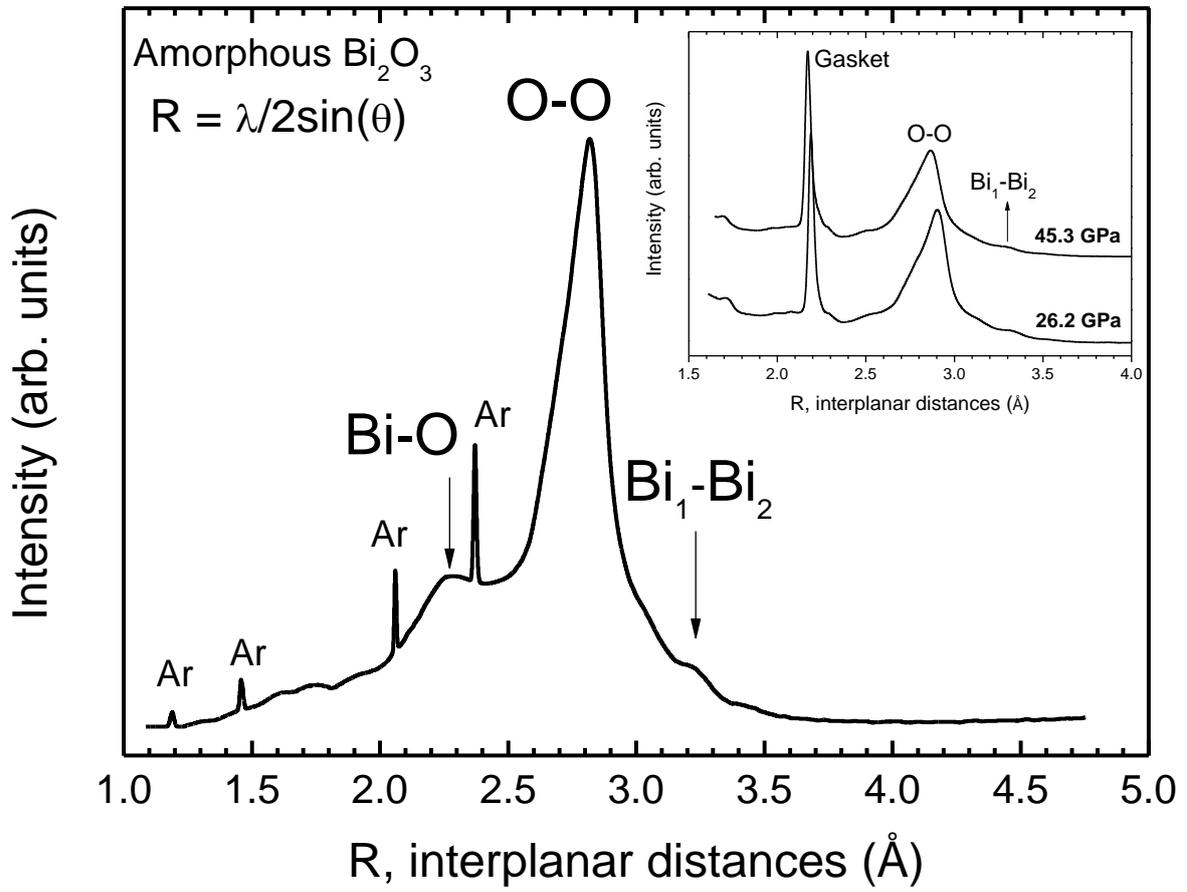

**Figure 8**

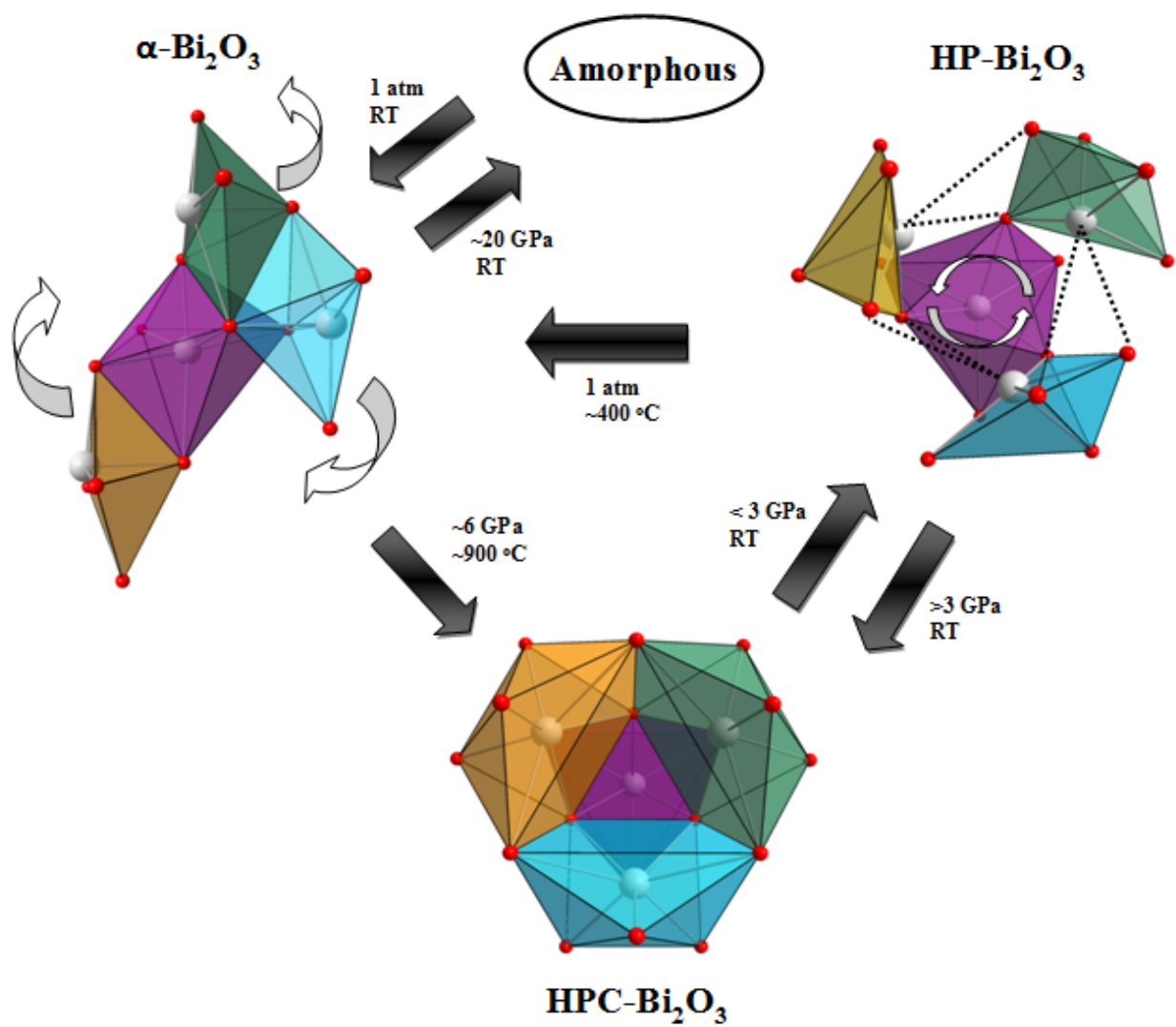
30